\documentclass[10pt,letterpaper]{article}

\usepackage{cogsci}

\cogscifinalcopy 

\usepackage{cmap}
\usepackage[T1]{fontenc}
\usepackage[american]{babel}
\usepackage{csquotes}
\usepackage{graphicx}
\usepackage{newtxtext,newtxmath}  

\usepackage[
  backend=biber,
  style=apa,
  natbib=true,
  annotation=false,
]{biblatex}
\usepackage{float} 

\usepackage[hidelinks]{hyperref}

\usepackage[labelfont=bf]{caption}
\usepackage{subcaption}
\newcommand\blfootnote[1]{%
  \begingroup
  \renewcommand\thefootnote{}\footnote{#1}%
  \addtocounter{footnote}{-1}%
  \endgroup
}

\title{Pragmatic Reasoning in Design}

\author{
  \vspace{1em}
  {\normalsize \bfseries Lance Ying$^{1,2}$, \quad William Van Uitert$^1$,\quad Tan Zhi-Xuan$^3$, \\ \bfseries Joshua B. Tenenbaum$^2$, \quad Samuel J. Gershman$^1$} \\ 
  \vspace{1em}
  {\normalsize\normalfont
    $^1$Harvard University  \quad $^2$Massachusetts Institute of Technology \quad  $^3$ National University of Singapore
  }
}

\begin{document}

\maketitle

\begin{abstract}
People can often understand and use novel artifacts after only a few interactions, suggesting that design choices communicate underlying affordances and causal structure. We propose a formal account of this process by framing cooperative, user-centered design as a cooperative game in which the user is the principal and the designer is an assistant. Inspired by prior work on pragmatic communication (e.g. RSA), our model treats a designer’s design decisions as communicative signals and predicts user judgments via recursive mentalizing: designers make design decisions to trade off informativeness about the artifact with efficiency, and users infer the true model of the artifact by inverting this cooperative designer model. We evaluate the model in a "design game" where designers place visually identical keys on trays to help a user infer which keys unlock which doors in grid-world layouts. We find that pragmatic designer and user models better match human judgments than non-mentalizing literal baselines.
\blfootnote{Published as a conference paper at CogSci 2026}

\end{abstract}

\section{Introduction}
People routinely learn to use unfamiliar artifacts after only a few interactions, suggesting that design choices provide cues about an artifact’s affordances and underlying causal structure. A central challenge for cognitive science and human-computer interaction (HCI) research is to explain how users form these mental models, and how designers can intentionally shape them.

Prior work in user-centered design emphasizes that designers should anticipate users' goals, limitations, and likely confusions when creating an artifact \citep{Norman2013}. Despite interests in the intersection between design and cognitive science, there has not been a formal account of this process, and it's unclear how users interpret and interact with artifacts that are intentionally designed for them.

\begin{figure*}[ht]
    \centering
    \begin{subfigure}[t]{0.32\textwidth}
        \centering
        \includegraphics[height=3.5cm]{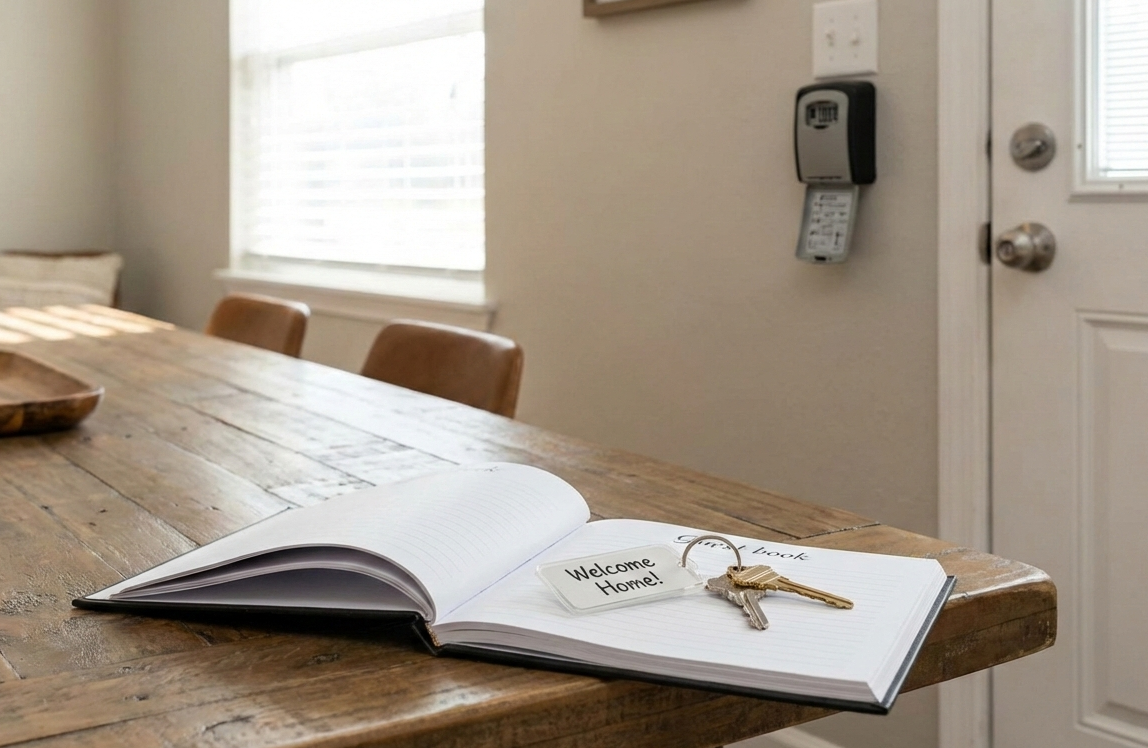}
        \caption{AirBnB key placement}
        \label{fig:design-airbnb}
    \end{subfigure}%
    \begin{subfigure}[t]{0.32\textwidth}
        \centering
        \includegraphics[height=3.5cm]{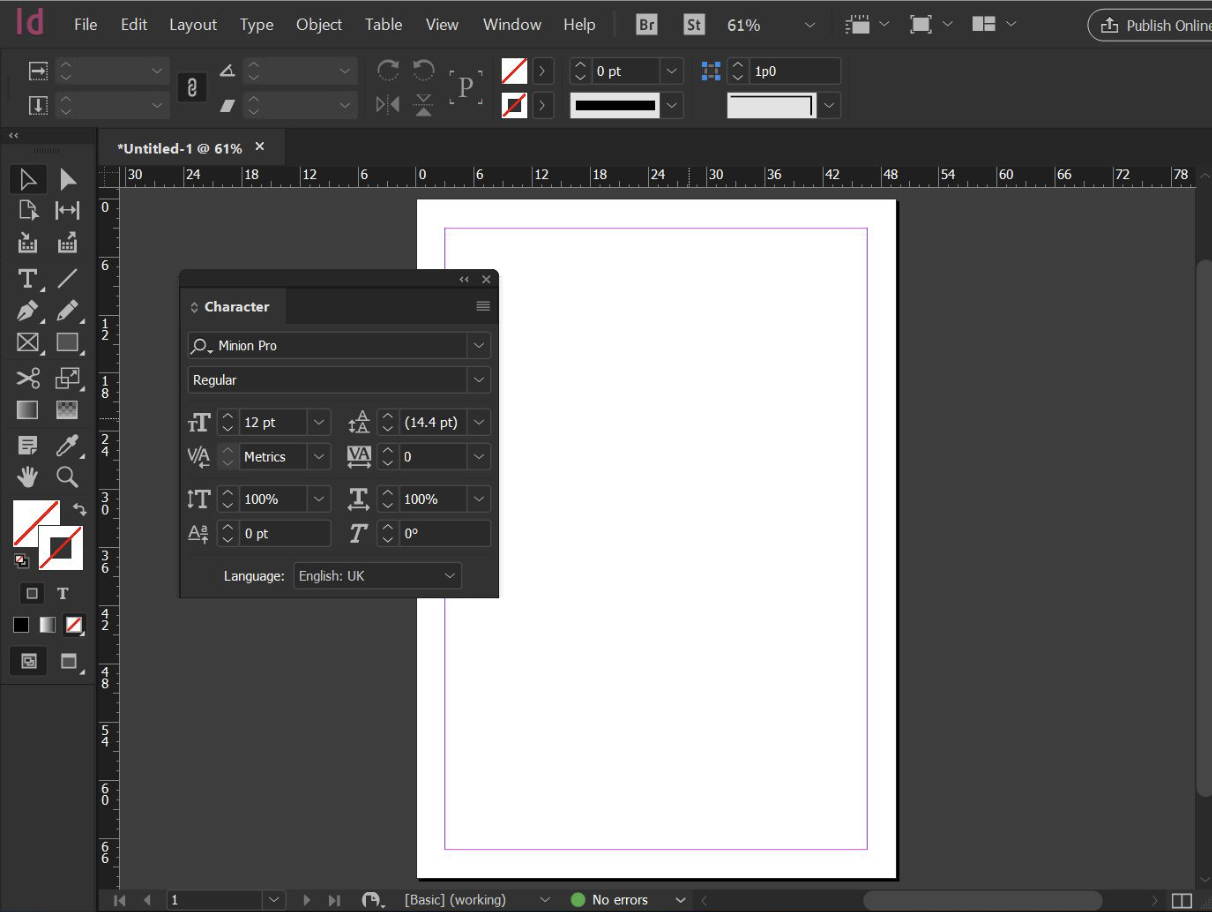}
        \caption{Adobe InDesign user interface.}
        \label{fig:design-adobe}
    \end{subfigure}%
    \hspace{2pt}
    \begin{subfigure}[t]{0.32\textwidth}
        \centering
        \includegraphics[height=3.5cm]{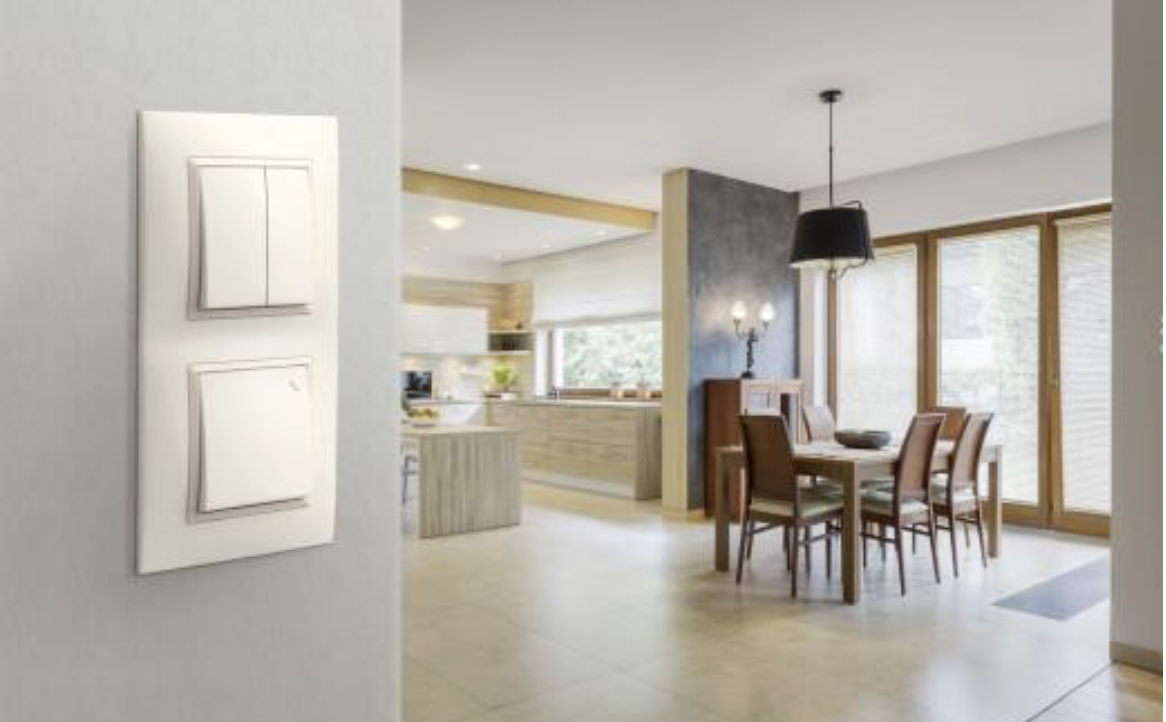}
        \caption{Room light switches}
        \label{fig:design-room}
    \end{subfigure}
    \caption{Example designs and pragmatic inference. \textbf{(a)} An AirBnB host intentionally left a key on a table with an open book. A pragmatic user can infer that the key must be related to the guest's booking. \textbf{(b)} Adobe InDesign UI which shows many buttons crowded in a toolbar. Each icon individually may offer limited information of their functionality, but by reasoning about how the icons are positioned, one may pragmatically infer additional information about the buttons. \textbf{(c)} Room design where switches on the wall may correspond to a particular appliance. A pragmatic reasoner may assume that the switches are designed to help one easily navigate a room.}
    \label{fig:design-illustration}
\end{figure*}

While previous computational accounts of pragmatic communication have focused on \emph{communicative intent recognition}---inferring an agent’s hidden communicative intent from observed actions, many design settings differ in a crucial way: goals are often \emph{common knowledge}. The designer knows the user’s goal and aims to help the user achieve it, and the user expects the designer to be helpful (see Figure \ref{fig:design-illustration} for an example).

This shifts the problem from ``What does the other agent want?'' to ``How does this artifact work?.'' Design often communicates a \emph{mental model} of an artifact---what actions are possible and what outcomes they produce---rather than a private intention. 

\begin{figure}[t!]
    \centering
    \includegraphics[width=\linewidth]{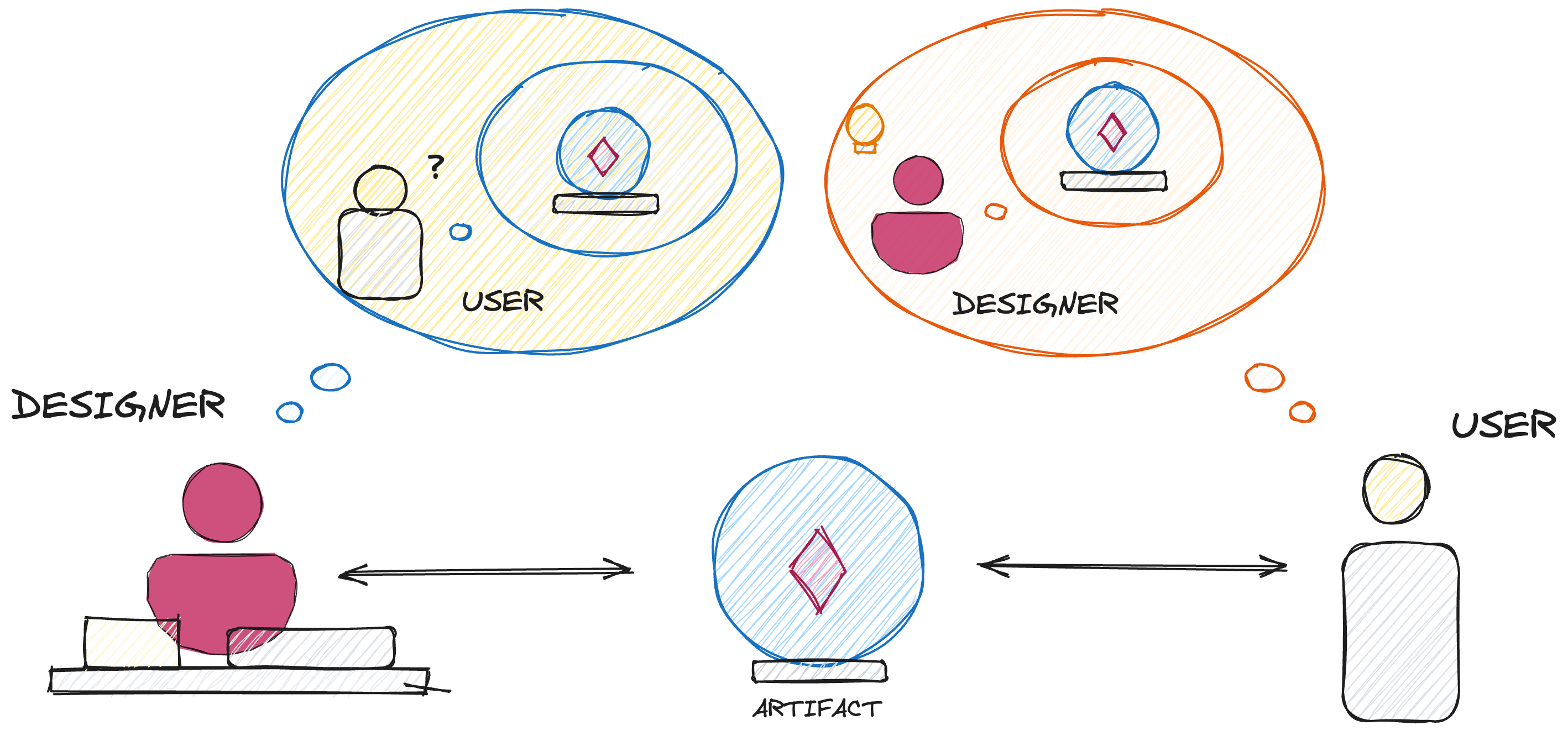}
    \caption{Illustration of recursive mentalizing in user-centered design, where the designer optimizes the artifact by simulating how the user may interpret the design, and the user leverage this process to infer the underlying model of the artifact by assuming it is the product of rational and intentional decision making by a cooperative mentalizing designer.}
    \label{fig:design_illustration}
\end{figure}

We propose a formal model that frames cooperative design as a \emph{design game} in which the \emph{user} is the principal and the \emph{designer} is an assistant. Inspired by pragmatic communication models such as the Rational Speech Act (RSA) framework \citep{GoodmanFrank2016}, we treat design choices as communicative actions: the designer recursively mentalizes about how the user will interpret available cues, and the user reasons about what a helpful designer would have built (Figure \ref{fig:design_illustration}).

We instantiate this framework in a room design game across a range of rooms and scenarios, where a knowledgeable designer places visually identical keys on trays in a partially filled grid-world rooms to help a naive user infer which keys unlock which doors. We develop pragmatic designer and pragmatic user models, along with non-mentalizing literal baselines.

We find that the pragmatic designer model predicts human design choices substantially better than the literal baseline. Together, these results support the claim that designers can reason about the users' goals and actions when designing an environment, while users can interpret novel designs by reasoning not only about perceptual cues, but also about a helpful designer’s communicative and cooperative intent.

\section{Related Work}
Our work connects research on: (i) design as communication and interpretability, (ii) pragmatic models of inference in language, and (iii) cooperative/assistance games in multi-agent interaction.

\paragraph{User-Centered Design}
A central idea in HCI is that successful artifacts are designed around users' goals, limitations, and expectations rather than around internal system structure. Classic accounts of user-centered design emphasize that designers should anticipate the user’s actions and likely confusions, making the correct actions easy to discover and execute while preventing or mitigating errors \citep{Norman2013,Nielsen1993, gibson2014theory}.

A practical implication is that designers often need to \emph{simulate} users: they evaluate candidate designs by predicting how a user would interpret available cues and what actions would follow. Methods such as cognitive walkthrough operationalize this idea by stepping through tasks from the perspective of a novice user and asking whether the interface makes the next action sufficiently salient \citep{Wharton1994}. Related cognitive modeling approaches (e.g., GOMS) similarly aim to predict usability and learning costs by approximating users' internal procedures \citep{CardMoranNewell1983}.


\paragraph{Pragmatic Communication}
We build on work in \emph{pragmatic communication} that treats communicative behavior as rational action shaped by an observer’s inference. In the Rational Speech Act (RSA) framework, speakers choose utterances to be informative with respect to a listener model, and listeners invert that process to infer intended meaning \citep{FrankGoodman2012,GoodmanFrank2016}. A closely related tradition models social inference as \emph{inverse planning}: observers infer hidden goals and intentions by assuming agents act (approximately) rationally to achieve them \citep{BakerSaxeTenenbaum2009,zhi2020online}. Recent work extends this idea to settings where agents intentionally shape others’ inferences \citep[``inverse inverse planning'';][]{ChandraEtAl2023}.

Work on \emph{pedagogy} also frames teaching as sequential decision making: teachers select demonstrations or interventions to guide a learner’s beliefs and policies, and learners interpret those demonstrations as communicative \citep{shafto2014rational,HoLittmanCushmanAusterweil2018,chen2024hierarchical}. More broadly, people can infer communicative intent from non-linguistic actions and artifacts, including how objects are placed in the environment \citep{RoykaEtAl2022,LopezBrauJaraEttinger2023}.


\paragraph{Assistance Games}
Assistance games and cooperative-inference formulations treat interaction as a team problem with asymmetric information: one agent knows the task or dynamics and must act so that a partner can succeed \citep{HadfieldMenell2016,Fisac2019,zhixuan2024pragmatic}. This perspective has been influential in multiagent systems, human-centered AI, and AI alignment, where agents are evaluated by how well they enable a human to achieve goals rather than by their own direct reward. Within this paradigm, human users know their own goals, but  (AI) assistants do not, so assistants must infer goals from user behavior, while users should legibly demonstrate their goals \citep{DraganSrinivasa2013} in order to receive helpful assistance.

A related line of work formalizes the idea that an observer may need to infer the intended objective from an imperfect specification; for example, inverse reward design treats a provided reward function as evidence about the designer’s true preferences under a model of bounded specification \citep{HadfieldMenell2017IRD}. In contrast to both assistance games and inverse reward design, we assume that the user's objective is known to the designer/assistant. Instead, it is the designer who possesses hidden knowledge about artifact function that the user must infer.


\section{Design as a Cooperative Game}

\begin{figure*}[ht]
    \centering
    \includegraphics[width=0.8\textwidth]{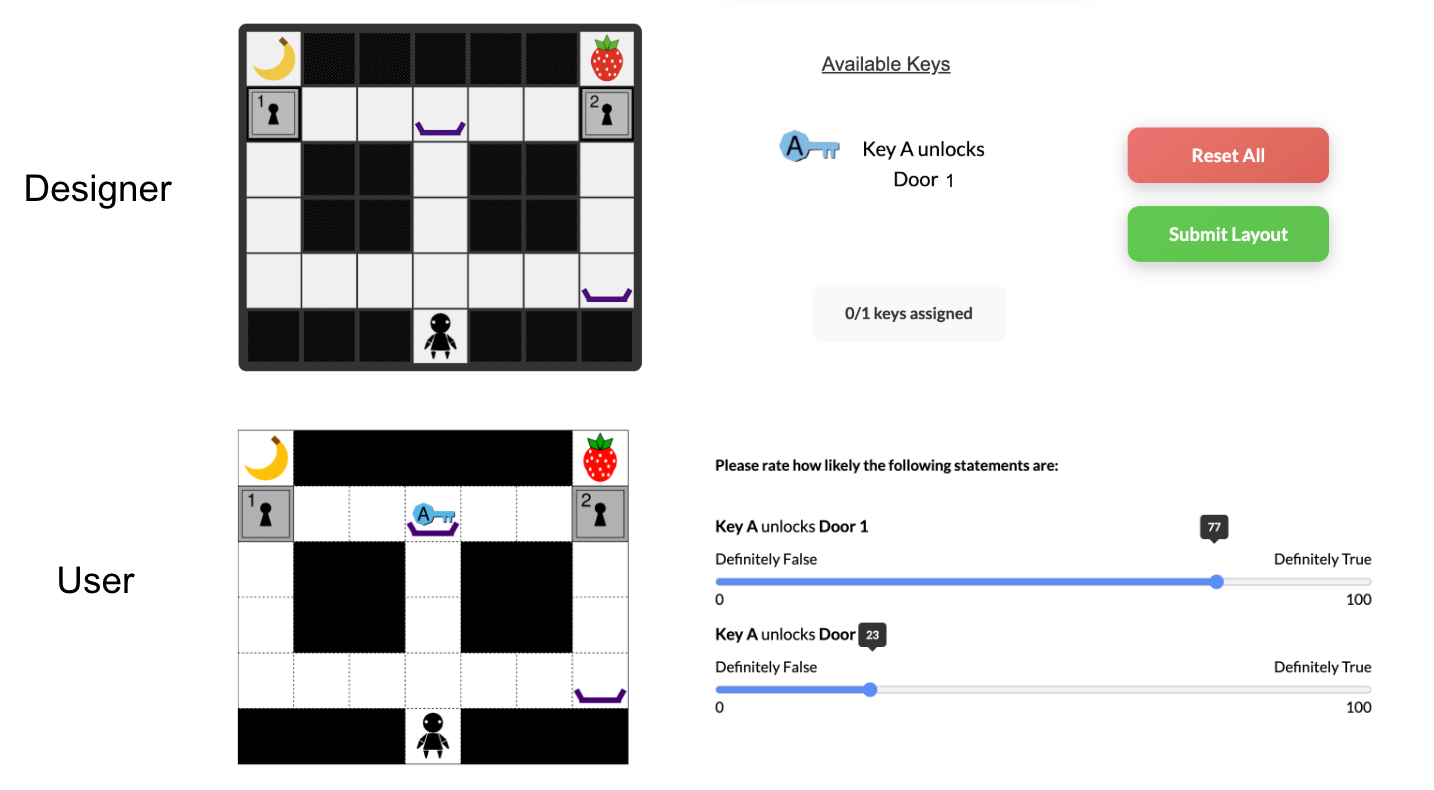}
    \caption{Experimental interface for the Room Design Game. (\textbf{Top}) \emph{The designer phase}: Human participants are recruited to act as designers in the game. They are provided a set of key(s) and the ground truth information on which door each key unlocks. The human participants are asked to drag keys into purple key trays in order to help the player (black agent). (\textbf{Bottom}) \emph{The user phase}: Human participants are recruited to act as player (black agent) in the game. They are told that the room was designed by another human to help them achieve the goal. The user agent is then asked to rate statements on what door they believe each key unlocks on a continuous scale.}
    \label{fig:interface}
\end{figure*}

Inspired by prior work on rational communication \citep{FrankGoodman2012} and assistance games \citep{HadfieldMenell2016}, we formalize cooperative design as a two-player cooperative game --- a \emph{design game} --- between a \textbf{designer} ($D$) and a \textbf{user} ($U$). The game takes place in an environment $\varepsilon$ consisting of three components: an initial state $s_0$, commonly-known causal structure $\varepsilon_\text{common}$, and hidden causal structure $\varepsilon_\text{hidden}$. Together, $\varepsilon_\text{common}$ and $\varepsilon_\text{hidden}$ determine how the user's actions affect the environment's state. Only the designer knows the full environment, including $\varepsilon_\text{hidden}$, while the user knows only $\varepsilon_\text{common}$ and must infer the hidden structure from the designed artifact.

The game proceeds in two phases. In the \emph{designer phase}, the designer modifies the initial state $s_0$ through a limited set of actions, producing a designed artifact $s_1$. We can think of $s_0$ as ``raw materials'' that the designer arranges into the artifact. In the \emph{user phase}, the user observes $s_1$ and takes actions toward their goal $g$. We assume the designer knows the user's goal. The designer's task is to construct the artifact so as to best enable the user to achieve their goal, which may involve communicating information about $\varepsilon_\text{hidden}$ through the design of $s_1$. Conversely, the user may need to infer the hidden structure from the design in order to achieve $g$ efficiently.

\subsection{The Room Design Game}

We instantiate this framework in the Room Design Game (Figure~\ref{fig:interface}). The environment is a gridworld containing fruits locked behind doors $d_1, d_2, \ldots$, with trays $t_1, t_2, \ldots$ at various locations. Each door is unlocked by exactly one key; keys are labeled but visually identical. The commonly-known structure $\varepsilon_\text{common}$ includes movement rules, that keys can be picked up, and that keys must be on trays. The hidden structure $\varepsilon_\text{hidden}$ is the key-door mapping $m^*$: for each key $k$, which door $d$ satisfies $\textsc{unlocks}(k, d)$. The user's goal $g$ is to obtain any of the fruits, which requires unlocking at least one door.

The initial state $s_0$ is the gridworld with empty trays. In the \emph{designer phase}, the designer places each key $k$ in some tray $t$---denoted $\textsc{placed}(k, t)$---producing the artifact $s_1$. The designer knows $m^*$ and knows the user lacks this information. In the \emph{user phase}, the user observes $s_1$ and must infer the hidden mapping. We measure users' beliefs by having them rate statements of the form ``key $k$ unlocks door $d$'' on a scale from 1 (definitely disagree) to 100 (definitely agree).

\begin{figure*}[ht]
    \centering
    \includegraphics[width=\textwidth]{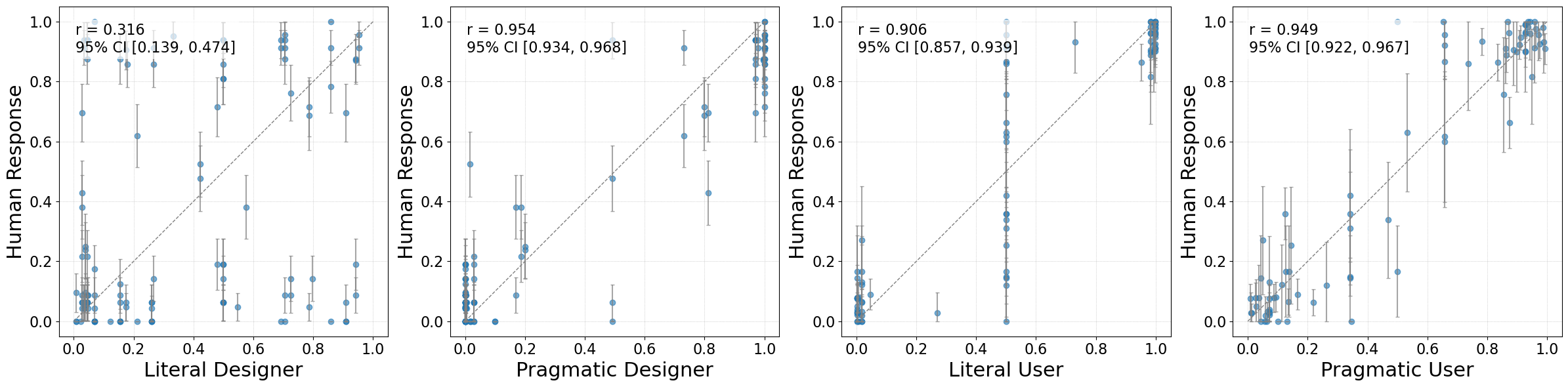}
    \caption{Model predictions compared to human judgments. Overall, we find that pragmatic models fit human judgments better than literal models in bother design and user experiments.}
    \label{fig:results}
\end{figure*}

\section{Computational Model}

The core insight of our model is that a pragmatic designer balances two considerations: the signal a placement provides about the hidden key-door mapping, and the efficiency of the placement for the user's task. A pragmatic user, in turn, assumes a pragmatic designer and inverts this model to infer the hidden mapping. We first describe the literal designer and user as baselines, then introduce the pragmatic agents.

\paragraph{Literal Designer ($D_1$).}
The literal designer places each key in a nearby tray (in terms of maze distance), treating design as efficiency optimization. Suppose $\textsc{unlocks}(k, d)$ holds in the true mapping $m^*$. The start position of the user is $s$. The utility of placing $k$ in tray $t$ is:
\begin{align}
    u_{D_1}\big(\textsc{placed}(k, t); m^* \big) = -\mathrm{Distance}(s, t, d).
\end{align}
where $\mathrm{Distance}(s, t, d)$ is the total distance of the user first going to tray $t$ and then to unlock door $d$. The designer's policy is a softmax over utilities:
\begin{align}
    P_{D_1}\big(\textsc{placed}(k, t) \mid m^* \big) \propto \exp\big(\beta \cdot u_{D_1}\big),
\end{align}
where $\beta \geq 0$ is an inverse temperature parameter.

\paragraph{Literal User ($U_1$).}
The literal user infers which door a key unlocks based on spatial proximity alone, without considering the designer's communicative intent. Given the observation that a key $k$ is on tray $t$, the literal user assumes $k$ is more likely to unlock nearby doors rather than far ones:
\begin{multline}
    P_{U_1}\big(\textsc{unlocks}(k, d) \mid \textsc{placed}(k, t) \big) \\ 
    \propto \exp\big(- \beta \cdot \mathrm{Distance}(s, t, d) \big).
\end{multline}

\paragraph{Pragmatic Designer ($D_2$).}
The pragmatic designer chooses placements that trade off informativeness --- how well a placement helps the user infer the true mapping --- against efficiency. Suppose key $k$ unlocks door $d$ in the true mapping $m^*$. Let $P_{U_1}(m \mid \textsc{placed}(k, t))$ be the literal user's posterior after observing placement in tray $t$. The designer's utility is:
\begin{align}
    u_{D_2}\big(\textsc{placed}(k, t); m^* \big) = 
    &-\alpha_1\, \mathrm{KL}\Big(P_{U_1}(m|\textsc{placed}(k, t)) \,\|\, \delta_{m^*}\Big) \notag \\
    &- \alpha_2\,\mathrm{Distance}(s, t, d),
\end{align}
where $\delta_{m^*}$ is the point mass at the true mapping, and $\alpha_1, \alpha_2 \geq 0$ weight the two terms. Their policy is:
\begin{align}
    P_{D_2}\big(\textsc{placed}(k, t) \mid m^*\big) \propto \exp\big(\beta \cdot u_{D_2}\big).
\end{align}

\paragraph{Pragmatic User ($U_2$).}
The pragmatic user assumes a $D_2$ designer and performs Bayesian inference to recover $m^*$:
\begin{multline}
    P_{U_2}(m^* \mid \textsc{placed}(k, t)) \propto \\
    \qquad \qquad P_{D_2}\big(\textsc{placed}(k, t) \mid m^*\big)\,P(m^*).
\end{multline}
That is, the pragmatic user inverts the designer's policy, reasoning about why a cooperative designer would have chosen the observed placement. We assume a uniform prior for $P(m^*)$.

\section{Experiment Design}

We designed 10 unique map layouts and generated 3 scenarios per map by varying the key-door mapping, for a total of 30 scenarios.

We recruited $60$ participants for the designer study (Mean age = 40.92, 37 male, 23 female) and $50$ participants for the user study (Mean age = 43.04, 15 male, 35 female). Because designs can vary across designers, we aggregate designs by taking the modal placement per trial and use that design as the stimulus for the user study. Each participant completes 10 trials in randomized order.

For computational models, we performed a grid search over all hyperparameters and retained the best fitting hyperparameters for analysis.

\begin{figure*}[ht]
    \centering
    \begin{subfigure}[b]{0.49\textwidth}
        \centering
        \includegraphics[width=\textwidth]{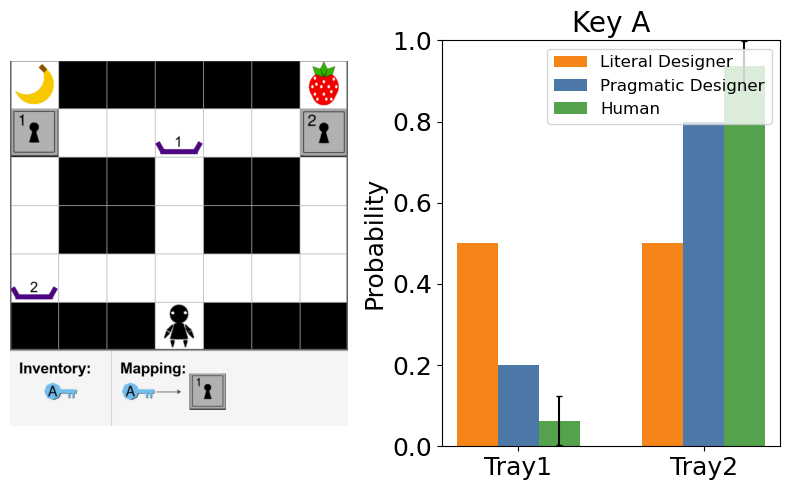}
    \end{subfigure}
    \hfill
    \begin{subfigure}[b]{0.49\textwidth}
        \centering
        \includegraphics[width=\textwidth]{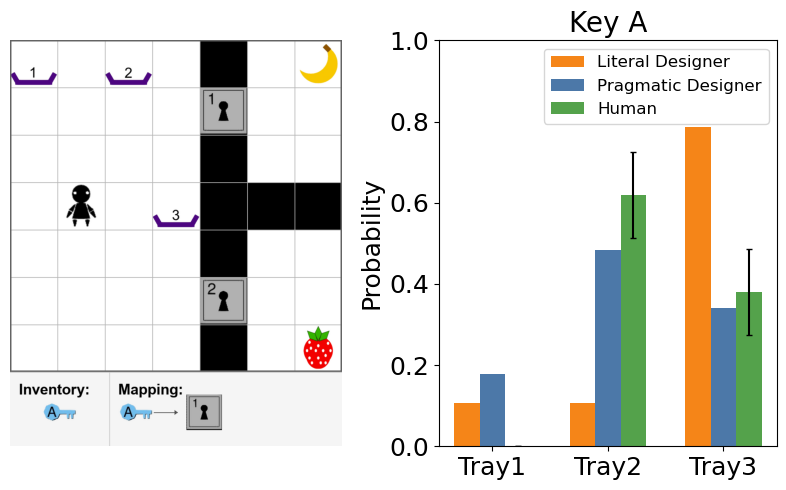}
    \end{subfigure}
    \caption{Qualitative designer examples comparing model predictions and human judgments. In each subplot, the map on the left shows the scenario, with the key available to the designer in the inventory and the ground truth mapping. Then the bar plot on the right shows the model prediction and averaged human response. We averaged all human participants' decisions to compute the probability of putting the key in each tray.}
    \label{fig:qualitative-designer}
\end{figure*}

\begin{figure*}[ht]
    \centering
    \begin{minipage}{0.49\textwidth}
        \centering
        \includegraphics[width=\textwidth]{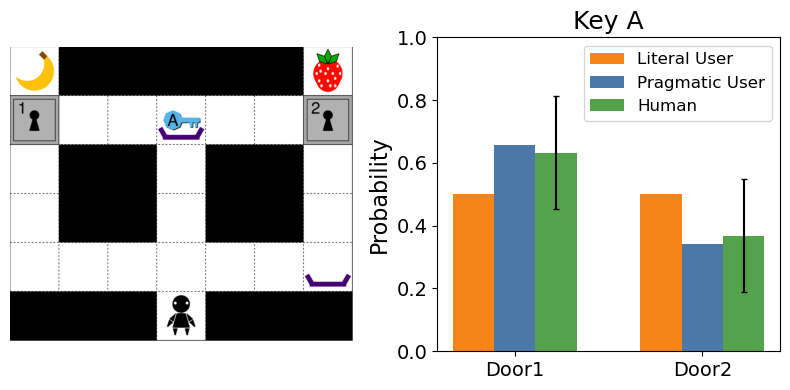}
    \end{minipage}
    \hfill
    \begin{minipage}{0.49\textwidth}
        \centering
        \includegraphics[width=\textwidth]{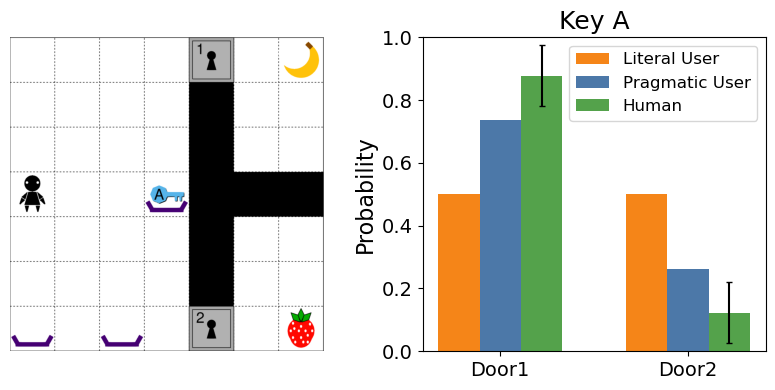}
    \end{minipage}
    \caption{Qualitative user examples comparing model predictions and human judgments. On the left of each subplot shows a map designed by a human participant in the designer game. On the right, each user sees the map and infer which door does the key unlock.}
    \label{fig:qualitative}
\end{figure*}

\subsection{Results}

Figure~\ref{fig:results} compares model predictions to human responses for both the designer and user tasks. Each point corresponds to a statement judgment in a given trial (normalized to $[0,1]$), and the dashed line indicates perfect agreement ($y=x$). Each dot represents an average probability measure for a given trial. For designer game, each dot represents the probability of assigning a key in a tray. If the trial has N keys and M trays, that would correspond to a total of N x M measures. For user game, each dot represents a statement rating from each trial. If the trial has N keys and M doors, that would correspond to a total of N x M ratings.

For the \textbf{designer phase}, the literal designer baseline shows only a modest correspondence with human placements ($r = 0.316$, 95\% CI $[0.139, 0.474]$). In contrast, the pragmatic designer model yields a substantially stronger fit ($r = 0.954$, 95\% CI $[0.934, 0.968]$), capturing designers’ tendency to make placements that are informative for the user rather than solely efficient.

For the \textbf{user phase}, both models predict human judgments well. The literal user achieves a strong fit ($r = 0.906$, 95\% CI $[0.857, 0.939]$), while the pragmatic user further improves performance ($r = 0.949$, 95\% CI $[0.922, 0.967]$), better matching the graded confidence users express when interpreting ambiguous placements.

Qualitatively, the literal user plot exhibits a pronounced vertical streak (many predictions concentrated at the rating of 0.5), even though human responses span a wide range. This pattern is consistent with the literal model producing discretized or tied posteriors when multiple doors are at similar distances (or when the softmax yields near-equal utilities), leading to the same predicted probabilities for many possible key-door mappings.

By comparison, the pragmatic user plot places substantially more mass near the diagonal, while also preserving the empirical tendency toward extreme judgments, indicating that both people and the pragmatic model often reach confident conclusions when placements are highly diagnostic.

To further illustrate these patterns, Figures~\ref{fig:qualitative-designer} and~\ref{fig:qualitative} show qualitative examples of human judgments alongside model predictions.

\subsection{Designer Phase Case Studies}

Figure~\ref{fig:qualitative-designer} illustrates two designer-trials. The left example shows a simple case where there are two purple trays. Tray 1 is equal distance between door 1 and door 2, and both trays are equal distance from door 1. The designer has Key A which unlocks door 1. We find that all human participants assigned the key to tray 1, which is predicted by the pragmatic designer model. This is because Tray 2 is more informative than Tray 1 to a user and thus have more communicative utility. In contrast, a literal designer is indifferent between Tray 1 and Tray 2 since they are equally efficient.

On the right, we show an example where designer needs to balance effiency and informativeness. In this example, the designer needs to allocate a key that unlocks door 1. Tray 3 is the most efficient option for door 1. A literal designer model therefore assign high probability of placing the key in Tray 3. However, Tray 3 is equal distance from both door 1 and 2, which makes it an option with low informativity. Tray 2, on the other hand, is slightly more inefficient but more informative. Both the pragmatic designer model and human participants show a lot more uncertainty between Tray 2 and Tray 3, indicating a tension between balancing efficiency and informativity in pragmatic design.

\subsection{User Phase Case Studies}

Figure~\ref{fig:qualitative} illustrates two representative user-trials where the same spatial arrangement can support different inferences depending on whether the user assumes the placement is merely efficient (literal) or intentionally informative (pragmatic).

In the left example, Key~A is placed in a tray adjacent to a corridor that separates the two doors. Under the literal user model, which relies primarily on spatial proximity, key~A is approximately equidistant from door~1 and door~2, producing near-indifference (roughly 0.5/0.5; orange bars). Human responses, however, strongly favor door~1 (green bar near 0.9), suggesting that users interpret the placement as an intentional cue rather than as a purely geometric signal. The pragmatic user partially captures this effect (blue bars), assigning higher probability to ``key~A unlocks door~1'' by reasoning that a cooperative designer would place a key so as to disambiguate the intended door. More generally, pragmatic inference can break symmetries in layouts where distance-only heuristics yield tied posteriors.

In the right example, the key is placed in the center tray, which is equal distance from both doors. Similar to the left example, the literal user interprets it to be equally likely to unlock door 1 and door 2. In contrast, both humans and the pragmatic user model assign higher probability to door 2, because if the key is supposed to unlock door 1, a pragmatic designer would have put the key in the left trays.

\subsection{Failure case}

While the pragmatic designer model achieved high correlation against human placement decisions, we also noticed a few outliers in the scatterplot (Figure \ref{fig:results}).

\begin{figure}
    \centering
    \includegraphics[width=\linewidth]{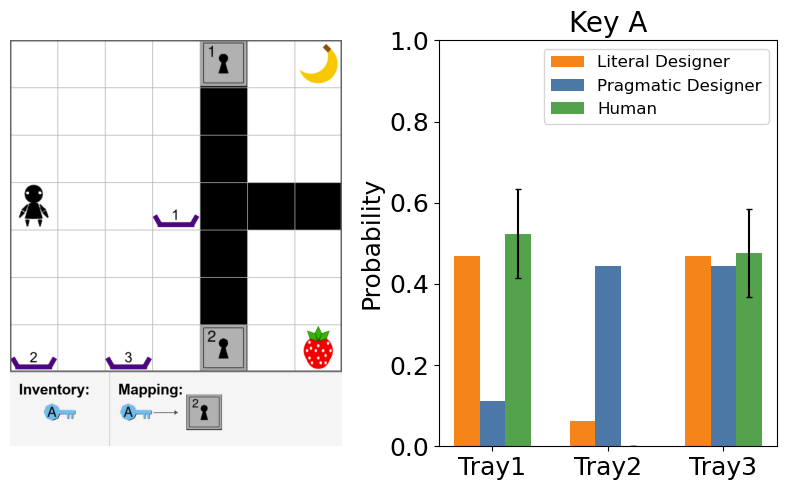}
    \caption{An example scenario where the pragmatic designer model does not match human judgment.}
    \label{fig:failure}
\end{figure}

The outliers are mainly due to one scenario shown in Figure \ref{fig:failure}. In this example, some human participants acted more like a literal designer, finding tray 1 to be an optimal place for the key that unlocks door 2. On the other hand, the pragmatic designer assigns equal utility for tray 2 and tray 3. This is because they are equally optimal and informative. We suspect that human participants may find tray 2 to be less optimal because it's farther away from door 2.

\section{Discussion and Conclusion}
Our results support the view that designers do more than optimizing for efficiency: they place cues in ways that anticipate a user’s downstream inferences. The pragmatic designer captures this strategic behavior substantially better than the literal baseline, suggesting that informativeness about hidden structure is an important component of design utility.

On the user side, the pragmatic user provides a consistent improvement over the literal distance-based model, even though the literal model already fits well. The qualitative structure of the scatterplots highlights a key difference: distance-based inference can collapse to tied or discretized predictions in symmetric layouts, whereas pragmatic inference smooths these ties by reasoning about why a cooperative designer would select particular placements. This aligns with the intuition that people interpret novel designs not only through perceptual cues, but also through assumptions about intentional, helpful design.

Our study is not without limitations. Our current formulation abstracts away several factors that likely matter in real artifacts, including heterogeneous user priors, learning over repeated interactions, and richer cost functions (e.g., physical effort, search, and error recovery). In addition, our evaluation uses a simplified grid-world task and a limited set of layouts, which may not capture the full complexity of everyday artifact understanding.

Future work may expand the space of communicative design actions beyond key placement (e.g., grouping, labeling, visual salience, or constraints that change the action set) and to study how multiple cues are integrated. We may consider more naturalistic design constraints such as physical laws. For instance, certain parts of the object may have physical affordances, which can affect the overall design (e.g., plane design needs to consider weight balancing for aerodynamics).

\printbibliography

\end{document}